\documentstyle[aps]{revtex}
\input epsf.sty          

\begin{document}
\draft

\title{The next to leading order effective potential in the 2+1
dimensional Nambu-Jona-Lasinio model at finite temperature.}

\author{F.P.~Esposito, 
I.A.~Shovkovy\thanks{On leave of absence from Bogolyubov Institute 
               for Theoretical Physics, Kiev 252143, Ukraine}, and 
L.C.R.~Wijewardhana}
\address{Physics Department, University of Cincinnati, 
         Cincinnati, OH 45221-0011, USA}
\date{\today}
\maketitle

\begin{abstract}
The finite temperature effective potential in the 2+1 dimensional
Nambu-Jona-Lasinio model is constructed up to the next to leading order
in the large $N$ expansion, where $N$ is the number of flavors in the
model. The distinctive feature of the analysis is an inclusion of an
additional scalar field, which allows us to circumvent the well known,
and otherwise unavoidable problem with the imaginary contribution to the
effective potential. In accordance with the Mermin-Wagner-Coleman
theorem, applied to the dimensionally reduced subsystem of the zero
Matsubara modes of the composite boson fields, the finite temperature
effective potential reveals a global minimum at the zero of the composite
order parameter. This allows us to conclude that the continuous global
symmetry of the NJL model is not broken for any arbitrarily small, finite
temperature.  
\end{abstract}  



\section{Introduction}

The Nambu-Jona-Lasinio (NJL) model \cite{NJL} plays a crucial role in
our understanding of quantum field theory. Despite its
non-renormalizability in 3+1 dimensions, the NJL model has already found
many applications in nuclear and high energy physics \cite{Klev,Volk}.

The 2+1 dimensional NJL model has also been studied in the large $N$
limit \cite{SemW,GatKRW}. This model is know to be non-renormalizable
when expanded in the four-fermi coupling. Nevertheless, it is
renormalizable in the $1/N$ perturbation theory \cite{Wil,RosWP1}. The
effective potential of this model has been studied at zero temperature 
up to the next to leading order in $1/N$ by carrying out the
Hubbard-Stratonovich transformation \cite{SemW,GatKRW}. When the
fluctuations of the composite scalars are included the effective
potential develops an imaginary part. At finite temperature this
complicates the analysis of the symmetry breaking and restoration. 

In this paper, we study the effective potential to the next to leading
order in the large $N$ expansion for the three-dimensional model at zero
and finite temperatures. The problem of obtaining the effective potential
is an old one. There exist well developed general methods for constructing
effective potentials for elementary \cite{ColW,Wei,Jac} as well as for
composite fields \cite{CJT} (see also \cite{GAC} where a finite
temperature version of the latter is discussed). 

These methods, however, cannot be applied to some specific models in a
straightforward manner. For example, should one try to construct the
effective  potential for a scalar theory with a double-well tree level
potential, he would find at the leading order an imaginary contribution
\cite{GatKRW,DolJ,Schn,Roo2}. The latter is due to a tachionic vacuum at
an intermediate stage in the calculations. In the scalar $\phi^4$ theory,
a method for dealing with this obstacle was found long ago
\cite{ColJP,Roo1,AboKS}. The way around the problem was the introduction
of an additional scalar field.

While studying the NJL model, one usually introduces some auxiliary
scalar fields that could be interpreted as composites built from fermions
\cite{SemW,RosWP1}. In terms of these new fields, the problem of
constructing the effective  potential is very much the same as in pure
scalar theories except that the propagators of the composite particles 
are much more complicated \cite{SemW,RosWP3}.

The leading order effective potential in the NJL model, which plays the
same role as a tree level potential in scalar theories, may also be
double-well in shape. In fact, this happens as soon as the four-fermi
coupling constant becomes larger than some critical value. Then, an
attempt to go beyond the leading order inevitably results in an imaginary
contribution to the effective potential due to the same reasons as in
the pure scalar case \cite{DolJ}.

Often there is no reason to doubt the leading order result, so this lack
of a straightforward method for calculating the next to leading order
corrections may seem to be of no interest. There are, however, several
instances when the knowledge of the next to leading  order terms becomes
of prime importance. This happens, for example,  when one studies the
finite temperature 2+1 dimensional NJL model which enjoys a continuous
global symmetry. The leading order calculation of the effective potential
indicates that the symmetry remains broken for a range of non-zero
temperatures. This result seem to contradict the predictions of
the  Mermin-Wagner-Coleman (MWC) theorem \cite{MWC}. The argument is 
well known. The infrared region of the system is dominated by the zero
Matsubara modes of the boson fields. The latter effectively live in a 1+1
dimensional space. Thus, by appealing to the above mentioned theorem, one
concludes that the symmetry breaking is forbidden \cite{Wit,MacPS}.

We find it surprising that there are no explicit calculations of the
next to leading order effective potential at finite temperature in the
NJL model in 2+1 dimensions to demonstrate that the symmetry is restored
for an arbitrarily small $T\neq 0$. In this paper we perform such
calculations and show that the global minimum of the potential does
indeed occur at the origin or, in other words, that the symmetry is not
broken at any finite temperature. The fluctuations of the composite
scalar particle, which is the NG boson at $T=0$, play a crucial role in
the symmetry restoration.

In some recent studies the finite temperature condensate
\cite{FloB,EbeNV} and the equation of state \cite{BCMPG} of NJL models 
have been computed to the next to leading order in $1/N$. It is
worthwhile to note that the problem of the imaginary contributions was
appreciated in \cite{BCMPG}. There it was noticed that the effective
potential of the NJL model (with a non-zero bare mass for fermions) is
real at the minimum field configuration. This is similar to the result
obtained by Dolan and Jackiw for theories with fundamental scalar fields 
\cite{DolJ}.  Even though the effective potential is real
at the minimum, the problem still exists in general.
In this paper we suggest a method that resolves the
imaginary part problem in the NJL model in 2+1 dimensions at finite
temperature (and even outside the minimum of the potential). Its
generalization may also work in the 3+1 dimensional case at finite 
temperature.

Finally, it seems that there is no clear understanding of what happens in
the 2+1 dimensional NJL model as the temperature approaches zero. Since
the MWC theorem is believed to work for arbitrarily small temperatures,
there is no symmetry breaking for any non-zero T. At T=0, the continuous
symmetry is broken. Then, a natural question arises. How does the
effective potential in the model behave as the temperature approaches
zero? We answer this question below.

\section{Zero temperature case}

Let us start with the analysis of the NJL model in 3-dimensional 
Euclidean space at zero temperature. To apply the large $N$ expansion
scheme, we consider the model containing $N$ identical flavors.  
The Lagrangian density reads 
\begin{equation}
L_{NJL}^{(0)} =\bar{\Psi} i\gamma_\mu \partial_\mu\Psi
- \frac{G}{2N} \left[(\bar{\Psi}\Psi)^2
+(\bar{\Psi}i\gamma_5\Psi)^2\right],
\label{NJLT=0}
\end{equation}
where we have chosen four-component spinors and the $4\times 4$ Dirac
$\gamma$-matrices are all antihermitian. By introducing (``integrating
in") the auxiliary boson fields  $\sigma\equiv(G/N)\bar{\Psi}\Psi$ and 
$\pi\equiv(G/N)\bar{\Psi}i\gamma_5\Psi$,  we obtain the equivalent 
model 
\begin{equation}
L_{aux}^{(0)} =
\bar{\Psi} \left(i\gamma_\mu \partial_\mu
-\sigma-i\gamma_5\pi\right)\Psi
+ \frac{N}{2G} \left(\sigma^2+\pi^2\right).
\label{auxT=0}
\end{equation}
The mentioned equivalence comes from the Euler-Lagrange equations of 
motion for the boson fields. As is easy to check, the latter Lagrangian 
density enjoys the same continuous chiral symmetry as that in 
Eq.~(\ref{NJLT=0}). The only difference is that the integrated in 
boson fields are also involved in symmetry transformations. 

After integrating out the fermions, we are left with the following
effective action for the bosons:
\begin{equation}
S_{scal}^{(0)} =N\left[\frac{1}{2G} \int d^3 x
\left(\sigma^2+\pi^2\right)
-Tr\ln\left(i\gamma_\mu \partial_\mu
-\sigma-i\gamma_5\pi\right)
\right].
\label{scalT=0}
\end{equation}
This action may be expanded in inverse powers of the cutoff
$\Lambda$ in the following manner,
\begin{equation}
S_{scal}^{(0)}  \simeq \frac{N}{\pi}\int d^3 x
\left[
-\frac{\Lambda}{2} \left(1-\frac{1}{g}\right)\rho^2
+\frac{\rho^3}{3}+\frac{\rho^4}{8\Lambda}
+O\left(\frac{\rho^6}{\Lambda^3}\right)
+L_{kin}^{(0)} (\sigma,\pi)
\right],
\label{AppScalT=0}
\end{equation}
where $\rho=\sqrt{\sigma^2+\pi^2}$ is chiral invariant, and the
dimensionless  coupling constant is defined as $g\equiv G\Lambda/\pi$. In
the kinetic part of the action, we keep only the terms that contain up
to two derivatives, i.~e., terms quadratic in momenta,
\begin{equation}
L_{kin}^{(0)} (\sigma,\pi)\simeq \frac{a_{\pi}^{(0)}}{2} 
\sum_{i=1}^{3}\left[ 
(\partial_{i}\sigma)^2 +(\partial_{i}\pi)^2 \right]
+\frac{a_{\sigma}^{(0)}-a_{\pi}^{(0)}}{2\rho^2} 
\sum_{i=1}^{3}\left(
\sigma\partial_{i}\sigma +\pi\partial_{i}\pi\right)^2.
\label{kinT=0}
\end{equation}
The straightforward calculation of the infrared asymptotic behavior 
of the boson propagators reveals the following expressions for the
kinetic coefficients \cite{GusM,BarL}
\begin{equation}
a_{\sigma}^{(0)}= \frac{1}{6\rho}, \qquad
a_{\pi}^{(0)} = \frac{1}{4\rho} .
\label{aT=0}
\end{equation}
Instead of using these infrared asymptotes, we replace them with constant
coefficients and argue that this would give a reasonable approximation in
the whole range of momenta. The reasoning goes as follows. The composite
particles in the far infrared region (large distances) should presumably
behave as almost point-like objects. Their  kinetic coefficients would be
given by the expressions as in Eq.~(\ref{aT=0}) with a specific value
of $\rho=\bar{\rho}$. As for  the ultraviolet region, it is known that
the leading asymptotic behavior does not depend on $\rho$. Therefore, we
take $a^{(0)}$'s independent of $\rho$ as a much better approximation
than that in Eq.~(\ref{aT=0}). Later, when we match the zero and the
finite temperature results, we will see that there is another advantage to
the above approximation.

Following the standard analysis applied for scalar theories
\cite{ColJP,AboKS}, we introduce an additional auxiliary scalar 
field $\chi$ into our bosonic action, 
\begin{equation}
S_{scal}^{(0)}  \simeq \frac{N}{\pi} \int d^3 x
\left(-\frac{\Lambda}{2}\chi^2+\frac{\chi}{2} 
\left(\rho^2-2\Lambda\mu\right)-\frac{\Lambda\mu^2}{2} 
+\frac{\rho^3}{3}
+L_{kin}^{(0)}(\sigma,\pi)
\right),
\label{chiT=0}
\end{equation}
where $\mu=\Lambda(g-1)/g$ is the characteristic scale of chiral 
symmetry breaking. Below we assume that $\mu\ll \Lambda$. As we 
mentioned in the Introduction, the model at hand is renormalizable in
the $1/N$ expansion. In spite of that, it is more convenient for us to
use the bare quantities throughout the paper without performing
renormalizations. That is sufficient for our purposes since we are
only interested in analyzing the long distance behavior of the theory,
which is captured quite well by a model with a large but finite physical
cutoff.

In calculations, we have to keep all contributions up to the order
$1/\Lambda$. However, it will be more convenient to keep exact results 
as they appear after integrations, without continually performing the
expansions. Obviously, this will not affect any of our results since all
the interesting features of the effective potential, as we shall see,
occur at values of $\sigma$ much smaller than the cutoff $\Lambda$. 

To construct the effective potential, we apply the standard technique
\cite{Jac} to the action in (\ref{chiT=0}). After performing rather
simple calculations up to the next to leading order of the large $N$
expansion, we obtain the following effective potential:
\begin{eqnarray}
V_{eff}^{(0)} (\sigma,\chi)&\simeq& 
\frac{N}{\pi}
\left(-\frac{\Lambda}{2}\chi^2
+\frac{\chi}{2} \left(\sigma^2-2\Lambda\mu\right)
-\frac{\Lambda\mu^2}{2}+\frac{\sigma^3}{3}\right)\nonumber\\
&&+\frac{1}{6\pi \left(a_{\sigma}^{(0)}\right)^{3/2} }\left[
\left(M_{\sigma}^{(0)}+a_{\sigma}^{(0)}\Lambda^2\right)^{3/2}
-\left(a_{\sigma}^{(0)}\Lambda^2\right)^{3/2}
-\left(M_{\sigma}^{(0)}\right)^{3/2}
\right]\nonumber\\
&&+\frac{1}{6\pi \left(a_{\pi}^{(0)}\right)^{3/2} }\left[
\left(M_{\pi}^{(0)}+a_{\pi}^{(0)}\Lambda^2\right)^{3/2}
-\left(a_{\pi}^{(0)}\Lambda^2\right)^{3/2}
-\left(M_{\pi}^{(0)}\right)^{3/2}
\right],
\label{EffchiT=0}
\end{eqnarray}
where the ``mass" parameters are
\begin{eqnarray}
M_{\sigma}^{(0)} &=&\chi+\frac{\sigma^2}{\Lambda}+2\sigma
\quad\mbox{and}
\label{MsigmaT=0}\\
M_{\pi}^{(0)} &=&\chi+\sigma.
\label{MpiT=0}
\end{eqnarray}
These parameters define the bosonic fluctuations. Since their propagators
do not have the canonical form, $M_{\sigma}^{(0)}$ and $M_{\pi}^{(0)}$
should not be confused with the sigma and pion masses.    

Finally, to get the effective action for the $\sigma$ field, we have to
eliminate the $\chi$ field by the saddle point equation
\begin{eqnarray}
\chi+\mu-\frac{\sigma^2}{2\Lambda}
-\frac{1}{4 N\Lambda \left(a_{\sigma}^{(0)}\right)^{3/2} }\left[
 \sqrt{M_{\sigma}^{(0)}+a_{\sigma}^{(0)}\Lambda^2} 
-\sqrt{M_{\sigma}^{(0)}}
\right]&&\nonumber\\
-\frac{1}{4 N \Lambda \left(a_{\pi}^{(0)}\right)^{3/2} }\left[
\sqrt{M_{\pi}^{(0)}+a_{\pi}^{(0)}\Lambda^2}-\sqrt{M_{\pi}^{(0)}}
\right]&=&0.
\label{SadleT=0}
\end{eqnarray}
We solve this equation numerically to obtain $\chi=\chi(\sigma)$. 
Then, we substitute it into Eq.~(\ref{EffchiT=0}) to get the effective
potential as a function of $\sigma$. The minimum of the obtained
effective potential satisfies the following gap equation,
\begin{eqnarray}
\chi\sigma 
+\sigma^2 +\frac{1+\sigma/\Lambda}
{2 N \left(a_{\sigma}^{(0)}\right)^{3/2} }\left[
 \sqrt{M_{\sigma}^{(0)}+a_{\sigma}^{(0)}\Lambda^2} 
-\sqrt{M_{\sigma}^{(0)}}
\right]&&\nonumber\\
+\frac{1}{4 N \left(a_{\pi}^{(0)}\right)^{3/2} }\left[
\sqrt{M_{\pi}^{(0)}+a_{\pi}^{(0)}\Lambda^2}-\sqrt{M_{\pi}^{(0)}}
\right]&=&0.
\label{GapT=0}
\end{eqnarray}

Prior to presenting the numerical results, let us analyze the
saddle-point equation for the $\chi$ field given in Eq.~(\ref{SadleT=0}).
It is easy to see that the left hand side of this equation is a
monotonically increasing function of $\chi$. We also observe that this
function takes real values only for $\chi\geq-\sigma$, and so, it is
bounded below with the minimum at $\chi=-\sigma$. If this minimum value
is positive, the left hand side remains positive in the whole range 
$\chi\geq-\sigma$, and the saddle point equation does not have a real
solution. Therefore, the existence of a real solution $\chi=\chi(\sigma)$ 
requires the following condition to be satisfied,
\begin{eqnarray}
\mu-\sigma-\frac{\sigma^2}{2\Lambda}
-\frac{1}{4 N\Lambda \left(a_{\sigma}^{(0)}\right)^{3/2} }\left[
 \sqrt{\frac{\sigma^2}{\Lambda}+\sigma+a_{\sigma}^{(0)}\Lambda^2} 
-\sqrt{\frac{\sigma^2}{\Lambda}+\sigma}
\right]-\frac{1}{4 N a_{\pi}^{(0)} }  &\leq&0.
\label{CondT=0}
\end{eqnarray}
This condition may be rewritten in a much more transparent way. Indeed,
if we choose a large enough value of $N$ the condition in
Eq.~(\ref{CondT=0}) does not hold for $\sigma=0$. By combining this with
the fact that the function in the left hand side of Eq.~(\ref{CondT=0})
decreases with increasing $\sigma$ (this is true everywhere except for a
small unimportant region of order $\mu^3/(N\Lambda)^2$ around
$\sigma=0$), we conclude that the above condition is equivalent to
$\sigma>\sigma_{cr}$, where  
\begin{equation}
\sigma_{cr}\approx \mu-\frac{a_{\sigma}^{(0)}+a_{\pi}^{(0)}}
{4Na_{\sigma}^{(0)}a_{\pi}^{(0)}}
-\frac{\mu^2}{2\Lambda}+O\left(\frac{1}{N\Lambda}\right).
\label{sigma_cr} 
\end{equation}

For any value of $\sigma$ greater than $\sigma_{cr}$, we can numerically
solve the saddle point equation (\ref{SadleT=0}). Then substituting
the solution $\chi=\chi(\sigma)$ into Eq.~(\ref{EffchiT=0}), we finally
obtain the effective potential as a function of $\sigma$ alone. Our
numerical analysis shows that the potential is a monotonically increasing
function in the region  $\sigma>\sigma_{cr}$ for a very wide range of
input parameters, which are the cutoff ($\Lambda$), the temperature
($T$), the number of flavors ($N$), the kinetic coefficients
($a_{\sigma}^{(0)}$, and $a_{\pi}^{(0)}$), and finally the scale
$\mu=\Lambda(g-1)/g$, which we put to 1 without loss of generality. In
Figure~1, we present the plot of the effective potential in the region
$\sigma>\sigma_{cr}$ obtained for a specific set of the parameters. 

Obviously, the straightforward extension of our calculations to the
region $0<\sigma<\sigma_{cr}$ would result in a non-zero imaginary part
in the effective potential. Such a situation looks rather disappointing.
It has been claimed, though, that the Maxwell construction has to be used
for the region $0<\sigma<\sigma_{cr}$ \cite{RosWP2}. The latter solution
of the problem  presumably corresponds to the flat potential in the
``forbidden" range of $\sigma$ as is plotted in Figure~1. We would like 
also to mention Ref.~\cite{MukS} where a strong argument in support of
the flat potential was given. This behavior of the potential is
interpreted as a signal for the phase with the broken symmetry. 

\section{Non zero temperature}

Now, let us study the NJL model at finite temperature. 
The partition function is given by the following path integral
\begin{equation}
Z=\int \left(\prod_{x,\tau} d \bar{\Psi} d \Psi\right)e^{-S_{NJL}},
\label{PF}
\end{equation}
where the functional measure is defined in a way consistent with the
standard anti-periodic boundary conditions for the fermion field. The
Euclidean ``action" that appears in the partition function is given by
\begin{equation} 
S_{NJL} =\int\limits_{0}^{\beta} d\tau \int d^2 x\left(
\bar{\Psi} i\gamma_\mu \partial_\mu\Psi
- \frac{G}{2N} \left[(\bar{\Psi}\Psi)^2
+(\bar{\Psi}i\gamma_5\Psi)^2\right]\right),
\label{NJL}
\end{equation}
where $\beta\equiv 1/T$ is the inverse temperature.

As in the case of zero temperature, we introduce two auxiliary  scalar
fields, so that the partition function is given by a path integral over
both the fermion and the boson fields. The corresponding action
reads
\begin{equation}
S_{aux} =\int\limits_{0}^{\beta} d\tau \int d^2 x\left[
\bar{\Psi} \left(i\gamma_\mu \partial_\mu
-\sigma-i\gamma_5\pi\right)\Psi
+ \frac{N}{2G} \left(\sigma^2+\pi^2\right)\right],
\label{aux}
\end{equation}

We are interested in the effective action for the  $\sigma$-field.
In order to obtain it, first, we calculate the expression for the
partition function in terms of the scalar fields alone by performing the
Gaussian integral over the fermions. As a result, we are left with the
following ``exact" action for the scalars,
\begin{equation}
S_{scal} =N\left[\frac{1}{2G} \int\limits_{0}^{\beta} d\tau \int d^2 x
\left(\sigma^2+\pi^2\right)
-Tr\ln\left(i\gamma^\mu \partial_\mu
-\sigma-i\gamma_5\pi\right)
\right],
\label{scal}
\end{equation}
and the partition function is given by the integral over the scalar
fields, 
\begin{equation}
Z=\int \left(\prod_{x,\tau} d \sigma d \pi \right)e^{-S_{scal}}.
\label{PFscal}
\end{equation}

Following closely our zero temperature calculation, we approximate
the action in Eq.~(\ref{scal}) by a truncated  derivative expansion
(which in momentum space corresponds to the expansion in powers
of momenta up to the second order). The approximate action is
\begin{eqnarray}
S_{scal} &\simeq& \frac{N}{\pi}\int\limits_{0}^{\beta} 
d\tau \int d^2 x \left[
-\frac{\Lambda}{2} \left(1-\frac{1}{g}\right)\rho^2
+\frac{\rho^3}{3}+\frac{\rho^4}{8\Lambda}
+O\left(\frac{\rho^6}{\Lambda^3}\right) 
\right.\nonumber\\
&&\left.
+2T\int_{0}^{\rho}dx x
\ln\left(1+\exp\left(-\frac{x}{T} \right)\right)
+L_{kin}(\sigma,\pi) \right],
\label{AppScal}
\end{eqnarray}
where we use the same notation as in the previous section. The 
kinetic term looks somewhat different from that at zero temperature,
\begin{eqnarray}
L_{kin}(\sigma,\pi)&\simeq& \frac{a_{\pi}}{2}
\sum_{i=1}^{2}\left[ 
(\partial_{i}\sigma)^2+(\partial_{i}\pi)^2\right]
+\frac{b_{\pi}}{2}\left[ 
(\partial_{\tau}\sigma)^2+(\partial_{\tau}\pi)^2\right]
\nonumber\\
&+&\frac{a_{\sigma}-a_{\pi}}{2\rho^2}
\sum_{i=1}^{2}\left( 
\sigma\partial_{i}\sigma+\pi\partial_{i}\pi\right)^2
+\frac{b_{\sigma}-b_{\pi}}{2\rho^2}\left( 
\sigma\partial_{\tau}\sigma+\pi\partial_{\tau}\pi\right)^2,
\label{kin}
\end{eqnarray}
where the coefficient $a$'s and $b$'s are now not equal. This is
due to the heat bath which always breaks Lorentz invariance. In
particular, the infrared asymptotic behavior of the kinetic term leads
to the following coefficients,
\begin{eqnarray}
a_{\sigma}&=&\frac{1}{6\rho} \left(
\tanh \left(\frac{\rho}{2T}\right)
+\frac{\rho}{4T\cosh^2(\rho/2T)}\right),
\label{asigma}\\
b_{\sigma}&=&\frac{1}{6\rho} \left(
\tanh \left(\frac{\rho}{2T}\right)
-\frac{\rho}{2T\cosh^2(\rho/2T)}
+\frac{\rho^2\tanh(\rho/2T)}
{2T^2\cosh^2(\rho/2T)}\right),
\label{bsigma}\\
a_{\pi}&=&\frac{1}{4\rho} 
\tanh \left(\frac{\rho}{2T}\right),
\quad\mbox{and}
\label{api}\\
b_{\pi}&=&\frac{1}{4\rho} \left(
\tanh \left(\frac{\rho}{2T}\right)
-\frac{\rho}{2T\cosh^2(\rho/2T)}\right).
\label{bpi}
\end{eqnarray}
As in the zero temperature case, we do not use these expressions later
in the calculation, but switch to the same approximation as before.
Namely, we assume that the $a$'s and $b$'s are independent of $\rho$.
Later we give arguments to substantiate this claim.

After ``integrating in" the auxiliary $\chi$ field, we arrive at the
following action,
\begin{eqnarray}
S_{scal} &\simeq& \frac{N}{\pi}\int\limits_{0}^{\beta} 
d\tau \int d^2 x \left[
-\frac{\Lambda}{2}\chi^2+\frac{\chi}{2} 
\left(\rho^2-2\Lambda\mu\right)-\frac{\Lambda\mu^2}{2} 
+\frac{\rho^3}{3} \right. \nonumber \\
&&\left.
+2T\int_{0}^{\rho}dx x\ln\left(1+\exp\left(-\frac{x}{T}
\right)\right)
+L_{kin}(\sigma,\pi) \right].
\label{chi}
\end{eqnarray}
This is used for the construction of the effective potential. As in
the case of zero temperature, we obtain the result of interest in the 
next to leading order of the large $N$ expansion:
\begin{eqnarray}
&&V_{eff}(\sigma,\chi)\simeq 
\frac{N}{\pi}
\left[-\frac{\Lambda}{2}\chi^2+\frac{\chi}{2} 
\left(\sigma^2-2\Lambda\mu\right)-\frac{\Lambda\mu^2}{2} 
+\frac{\sigma^3}{3}
+2T\int_{0}^{\sigma}dx x\ln\left(1+\exp\left(-\frac{x}{T}
\right)\right)\right]\nonumber\\
&&
+T\sum_{n=-\infty}^{\infty}\int\frac{d^2p}{(2\pi)^2}
\ln(M_{\sigma}+a_{\sigma}p^2+b_{\sigma}\Omega_{n}^2)
+T\sum_{n=-\infty}^{\infty}\int\frac{d^2p}{(2\pi)^2}
\ln(M_{\pi}+a_{\pi}p^2+b_{\pi}\Omega_{n}^2),
\label{Effchi}
\end{eqnarray}
where we define
\begin{eqnarray}
M_{\sigma}&=&\chi+\frac{\sigma^2}{\Lambda} 
+\frac{2\sigma}{1+\exp(-\sigma/T)}
+2T\ln\left[1+\exp(-\sigma/T)\right]
\quad\mbox{and}
\label{Msigma}\\
M_{\pi}&=&\chi+\sigma
+2T\ln\left[1+\exp(-\sigma/T)\right].
\label{Mpi}
\end{eqnarray}

After carrying out the summation over Matsubara frequencies and 
the integration over momenta (with the sharp cutoff at $\Lambda$), 
we arrive at the result
\begin{eqnarray}
&&V_{eff}(\sigma,\chi)\simeq\frac{N}{\pi}
\left[-\frac{\Lambda}{2}\chi^2+\frac{\chi}{2} 
\left(\sigma^2-2\Lambda\mu\right)-\frac{\Lambda\mu^2}{2} 
+\frac{\sigma^3}{3}
+2T\int_{0}^{\sigma}dx x\ln\left(1+\exp\left(-\frac{x}{T}
\right)\right)\right]\nonumber\\
&&+\frac{1}{6\pi a_{\sigma} \sqrt{b_{\sigma}} }\left[
\left(M_{\sigma} +a_{\sigma} \Lambda^2\right)^{3/2}
-\left(a_{\sigma} \Lambda^2\right)^{3/2}
-\left(M_{\sigma} \right)^{3/2}
\right]\nonumber\\
&&+\frac{1}{6\pi a_{\pi}\sqrt{b_{\pi}} }\left[
\left(M_{\pi} +a_{\pi} \Lambda^2\right)^{3/2}
-\left(a_{\pi} \Lambda^2\right)^{3/2}
-\left(M_{\pi} \right)^{3/2}
\right]\nonumber\\
&&-\frac{T}{2\pi a_{\sigma} }
\int_{0}^{M_{\sigma}}dx \ln\left(1-\exp\left(
-\frac{\sqrt{x}}{T\sqrt{b_{\sigma}}}
\right)\right)
-\frac{T}{2\pi a_{\pi} }
\int_{0}^{M_{\pi}}dx \ln\left(1-\exp\left(
-\frac{\sqrt{x}}{T\sqrt{b_{\pi}}}
\right)\right).
\label{Eff}
\end{eqnarray}
Despite the fact that the calculations were simplified due to the
nontrivial assumption on the form of the kinetic part of the boson
action, we argue that the result should be reliable. Indeed, the
essential details of both the infrared and the ultraviolet regions seem
to be captured if the kinetic coefficients are chosen to be independent
of $\sigma$. 

The only property that we require of the boson propagators in the
ultraviolet region is a soft dependence on the field $\sigma$. On the one
hand, this resembles the known asymptotic behavior of the composite boson
propagators \cite{SemW,RosWP3}, and, on the other, this allows us to
avoid the problem of an unphysical sensitivity on the details of the
ultraviolet region. 

As for the infrared behavior of the propagators, the argument is somewhat
trickier. We have to distinguish  between small and large values of the
field $\sigma$. Obviously, the most interesting region lies close to the
origin, since it is there where one would expect the essential changes in
the effective potential when the chiral symmetry is restored. So, suppose
that we consider $\sigma\alt T$. Then for $|\vec{p}|\alt T$, the  kinetic
coefficients approach constants of order $1/T$. Since only the zero
Matsubara mode is relevant in this case, we can also ignore the fact that
$b_{\sigma,\pi}\to 0$. On the other hand, if we consider the region
$\sigma\agt T$, then, strictly speaking, we have to take into account any
dependence of the $a$'s and $b$'s on the value of the field $\sigma$.
Fortunately, there is no reason to believe that a mild dependence of the
propagators on $\sigma$ in this region could change any of our
qualitative results. 

Now, let us mention that the na\"{\i}ve limit $T\to 0$ of our finite
temperature potential given in Eq.~(\ref{Eff}) coincides with the
effective potential obtained in the previous section for $T=0$ if we
assume that $a_{\sigma}, b_{\sigma}\to a_{\sigma}^{(0)}$ and $a_{\pi},
b_{\pi}\to a_{\pi}^{(0)}$. 

To eliminate the auxiliary field $\chi$, we have to solve the saddle 
point equation, as was done in the case of zero temperature. This saddle
point equation is obtained by taking the derivative of the effective
potential (\ref{Eff}) with respect to $\chi$:
\begin{eqnarray}
\chi&+&\mu-\frac{\sigma^2}{2\Lambda}
-\frac{1}{4 N\Lambda a_{\sigma} \sqrt{b_{\sigma}} }\left[
\sqrt{M_{\sigma} +a_{\sigma} \Lambda^2}
-\sqrt{M_{\sigma}}\right]
-\frac{1}{4 N\Lambda  a_{\pi}\sqrt{b_{\pi}} }\left[
\sqrt{M_{\pi} +a_{\pi} \Lambda^2}
-\sqrt{M_{\pi}}\right]\nonumber\\
&+&\frac{T}{2  N\Lambda a_{\sigma} } \ln\left(1-\exp\left(
-\frac{\sqrt{M_{\sigma}}}{T\sqrt{b_{\sigma}}}
\right)\right)
+\frac{T}{2 N\Lambda  a_{\pi} }\ln\left(1-\exp\left(
-\frac{\sqrt{M_{\pi}}}{T\sqrt{b_{\pi}}}
\right)\right)=0.
\label{Sadle}
\end{eqnarray}
Let us analyze this equation. As in the case of $T=0$, the left hand side
of Eq.~(\ref{Sadle}) is a monotonically increasing function of $\chi$.
Unlike the former, though, the left hand side of eq.~(\ref{Sadle}) is not
bounded below, but approaches negative infinity as  $\chi\to -\sigma 
-2T\ln[1+\exp(-\sigma/T)]$, so that $M_{\pi}\to 0$. This simple property
has far reaching consequences. The most important among them is that
the saddle point equation has a real solution for all values of 
$\sigma$ ! This is quite remarkable if we notice that Eq.~(\ref{Sadle}) 
in the ``na\"{\i}ve" limit $T\to 0$ coincides with its counterpart that
was obtained for $T=0$.
 
The saddle point equation for $T=0$, as we saw, does not have a solution
for $0<\sigma<\sigma_{cr}$, while the finite temperature equation allows
a real solution for all values of $\sigma$. How could this happen? The
answer is quite simple, and involves the way the limit $T\to 0$ is taken.
In the ``na\"{\i}ve" limit, we assume that $M_{\pi}$ and $M_{\sigma}$
become much larger than the value of $T$ as we approach zero temperature.
On the other hand, at any small but finite temperature, we find that the
solution to the saddle point equation requires that $M_{\pi} \sim T^2
b_{\pi}\exp(-N\Lambda f(\sigma)/T)$, where $f(\sigma)$ is a positive
function of $\sigma$ for $0<\sigma<\sigma_{cr}(T)$.
 
Therefore, we observe that the $T=0$ quantum field theory corresponds to
the ``na\"{\i}ve" limit when $T\to 0$ is taken before solving the saddle
point equation. We get a very different result by taking the limit in the
other order, i.e., solving the saddle point equation first and then taking
the value of $T$ to zero. 

The finite temperature gap equation is given by
\begin{eqnarray}
\chi\sigma 
+\sigma^2
+\frac{\partial_{\sigma}(M_{\sigma})}{4N a_{\sigma} 
\sqrt{b_{\sigma}} }\left[
\sqrt{M_{\sigma} +a_{\sigma} \Lambda^2}
-\sqrt{M_{\sigma}}\right]
+\frac{\partial_{\sigma}(M_{\pi})}{4N a_{\pi}\sqrt{b_{\pi}} }\left[
\sqrt{M_{\pi} +a_{\pi} \Lambda^2}
-\sqrt{M_{\pi}}\right]&&\nonumber\\
-\frac{T}{2N a_{\sigma} } \partial_{\sigma}(M_{\sigma})
\ln\left(1-\exp\left(
-\frac{\sqrt{M_{\sigma}}}{T\sqrt{b_{\sigma}}}
\right)\right)
-\frac{T}{2N a_{\pi} } \partial_{\sigma}(M_{\pi})
\ln\left(1-\exp\left(
-\frac{\sqrt{M_{\pi}}}{T\sqrt{b_{\pi}}}
\right)\right)&=&0,
\label{Gap}
\end{eqnarray}
and admits the trivial solution $\sigma=0$. For $T=0$, this was not the
case. Actually, the existence of the trivial solution is a necessary
condition for concluding that the symmetry is restored, but not a
sufficient condition. Below, while presenting our numerical results, we
will see that this $\sigma=0$ solution is the only solution to the gap
equation, and that it corresponds to the global minimum of the effective
potential. 

Our numerical analysis begins with a solution of the saddle point
equation (\ref{Sadle}). While solving it, we notice that there are two
qualitatively different regions of the field $\sigma$. For small enough
values of the field $\sigma<\sigma_{cr}(T)$, where the $T=0$  equation
would not have a solution, we can give a very good analytical
estimate for the finite temperature solution,
\begin{equation}
\chi=-\sigma-2T\ln\left[1+\exp(-\sigma/T)\right]+\tilde{M}_{\pi},
\label{chi(s)}
\end{equation}
where
\begin{equation}
\tilde{M}_{\pi}\sim T^2 b_{\pi}
\exp\left(-\frac{N\Lambda}{T}f(\sigma)\right),
\label{mpi}
\end{equation}
and where $f(\sigma)$ is a positive function of $\sigma$ for
$0<\sigma<\sigma_{cr}(T)$. After $\sigma$ reaches the value
$\sigma_{cr}(T)$, the  solution $\chi(\sigma)$ sharply changes its
behavior and starts to follow closely the solution found for $T=0$. The
results of our numerical solution to the saddle point equation for four
different temperatures are plotted in Figure~2.

By substituting the solution $\chi=\chi(\sigma)$ of the saddle point
equation into Eq.~(\ref{Eff}), we  obtain the effective potential as a
function of $\sigma$. The corresponding plots for the four different
temperatures are given in Figure~3. As we see, the effective potential
reveals the behavior which was expected: its global minimum lies at the
origin. The latter means that the chiral symmetry is not broken.

By comparing the zero and finite but small temperature effective
potentials, we find that they follow very closely each other in the range
of large enough values of the field $\sigma>\sigma_{cr}$.

Another property of the obtained effective potential which is worth
noting is its convexity. This is a feature that one expects for the
effective potential based on general arguments \cite{MukS,Fukuda,ORaiWY}.
To see that our effective potential is indeed convex, we have to show
that its derivative is non-negative in the whole range of field $\sigma$.
For this purpose, in Figure~4, we plot the left hand side of the gap
equation (\ref{Gap}). Up to an unessential factor, it is equal to the
derivative of the potential as a function of $\sigma$. As we see, the
latter is everywhere non-negative.

Before concluding this section, we have to note that all the qualitative
results, such as the existence of one global minimum at $\sigma=0$, the
convex character of the potential, and the two-branched character of the
solution to the saddle point equation, persist for a rather wide choice
of the input parameters.

\section{Conclusion}

In this paper we have studied the effective potential in the
three-dimensional  NJL model at finite and zero temperatures up to the
next to leading order in the large $N$ expansion. The distinctive feature
of our analysis is the introduction of an additional scalar field. This
allows us to  circumvent the well known, and otherwise unavoidable,
problem of the imaginary contributions to the finite temperature
effective potential.   

In agreement with the MWC theorem, the analysis of the effective
potential in the three-dimensional NJL model at finite temperature does
not reveal breaking of the continuous chiral symmetry. This remains true
for an arbitrarily small value of the temperature. The leading order
effective potential that indicates symmetry breaking changes drastically
in the next to leading order. The essential role is played by the
pion-like particle, which would have become the NG boson were the
symmetry broken. 

We present the numerical results that clarify and confirm our analytical
findings in Figures~1-4. The zero temperature effective potential is
given in Figure~1. Since it is not defined in the range of fields
$0<\sigma<\sigma_{cr}$, we extended it with the Maxwell construction to
the forbidden region. More rigorous arguments in support of such an
extension can be found in \cite{MukS}. Figures ~2 through 4 present the
results at finite temperature.

By comparing our analytical and numerical results for both $T=0$ and
$T\neq 0$, we come to the conclusion that the zero temperature field
theory can be defined in two different ways. One definition comes from
the standard $T=0$ quantum field theory. The other is obtained by the
limiting procedure $T\to 0$ from the Matsubara imaginary time formalism.

\section*{Acknowledgments} 

We would like to thank G.~Semenoff and P.~Suranyi for useful
discussions, and E.~Gorbar for valuable comments. This work was 
supported in part by the U.S. Department of Energy  Grant
\#DE-FG02-84ER40153. 


\begin{figure}
\epsfbox{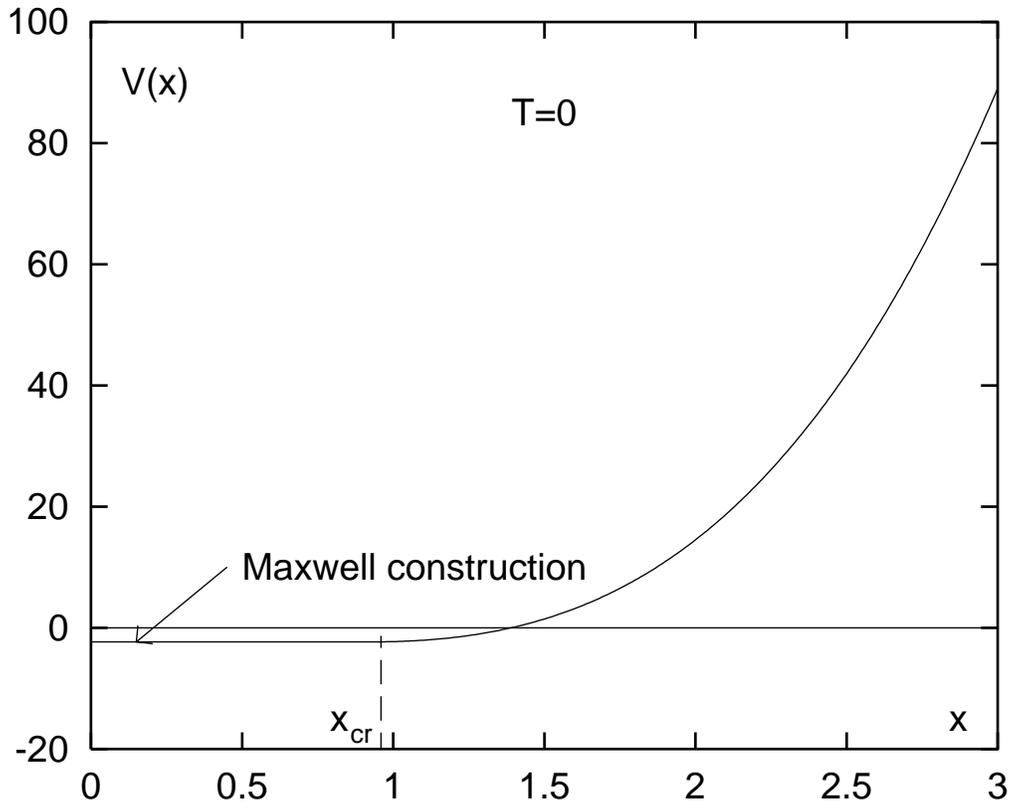}
\caption{The effective potential at zero temperature extended by the
Maxwell construction in the region of small values of the field. The 
results given in units of $\mu$: $V(x)=V_{eff}(x)/\mu^3$, $x=\sigma/\mu$.
The input parameters are  $\Lambda=10\mu$, $a_{\sigma}^{(0)}
=a_{\pi}^{(0)}=5/\mu$, and $N=50$.} 
\label{Figure1}
\end{figure}

\begin{figure}
\epsfbox{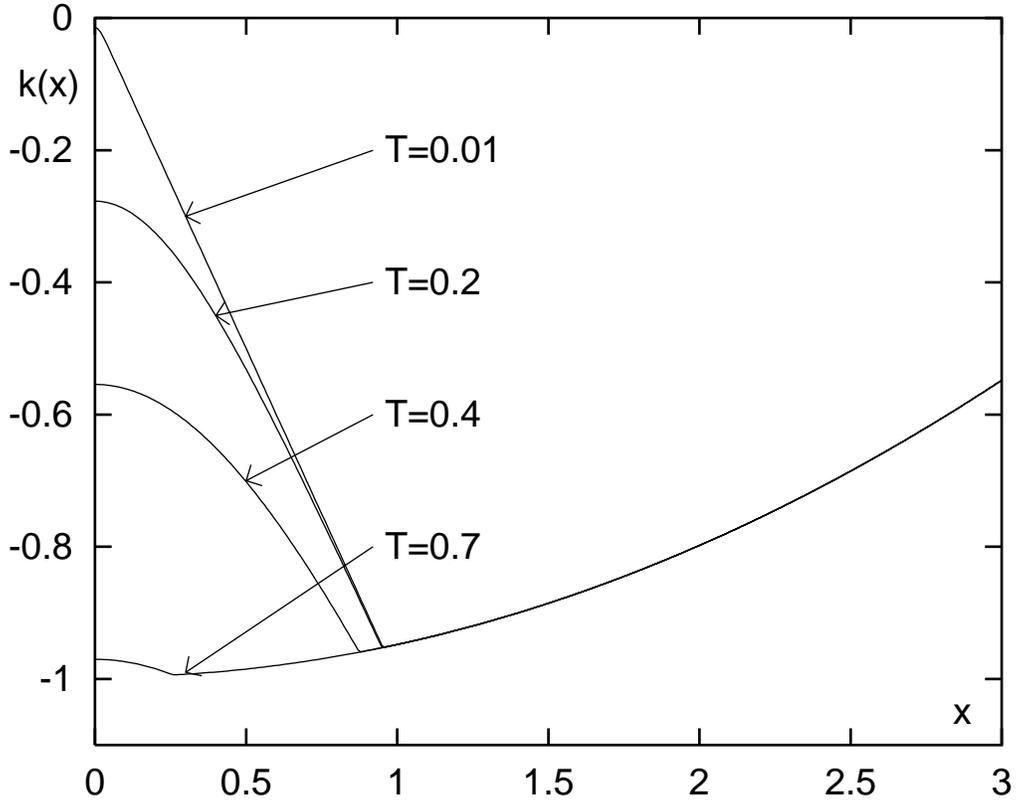}
\caption{The finite temperature solution to the saddle point equation 
given in units of $\mu$: $k(x)=\chi(x)/\mu$, $x=\sigma/\mu$. The input 
parameters are $\Lambda=10\mu$, $a_{\sigma}^{(0)}=a_{\pi}^{(0)}=5/\mu$, 
and $N=50$.}
\label{Figure2}
\end{figure}

\begin{figure}
\epsfbox{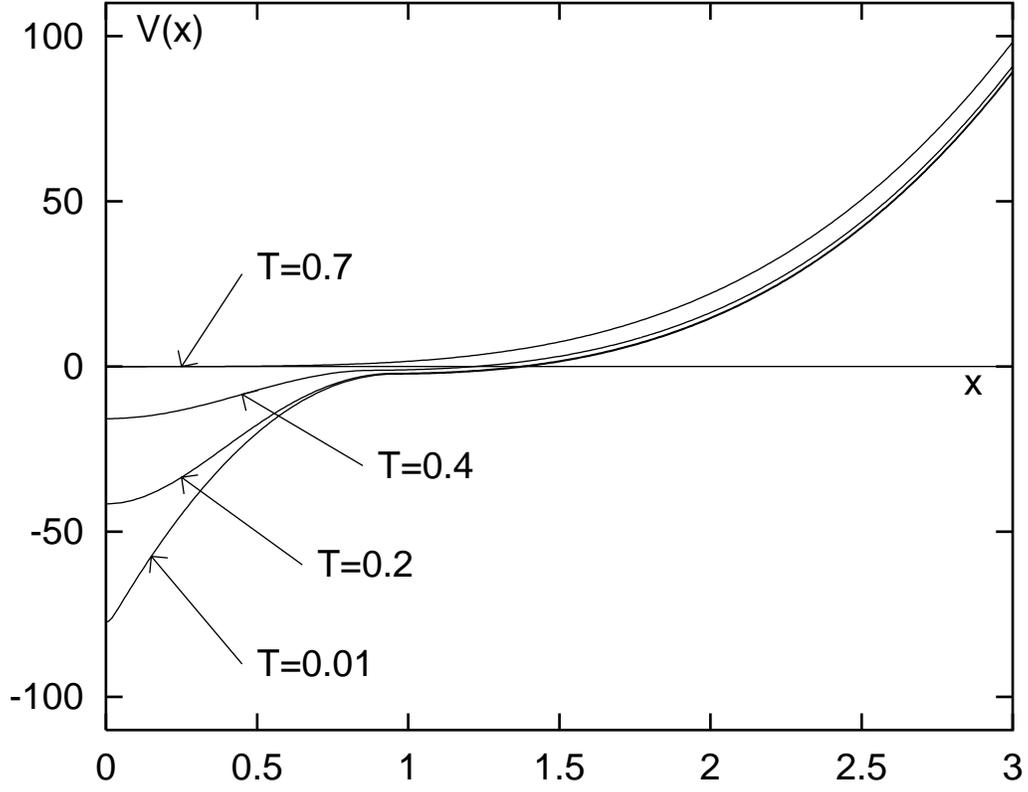}
\caption{The effective potential at non-zero temperatures given in 
units of $\mu$: $V(x)=V_{eff}(x)/\mu^3$, $x=\sigma/\mu$. The input 
parameters are  $\Lambda=10\mu$, $a_{\sigma}^{(0)}=a_{\pi}^{(0)}=5/\mu$, 
and $N=50$.}
\label{Figure3}
\end{figure}

\begin{figure}
\epsfbox{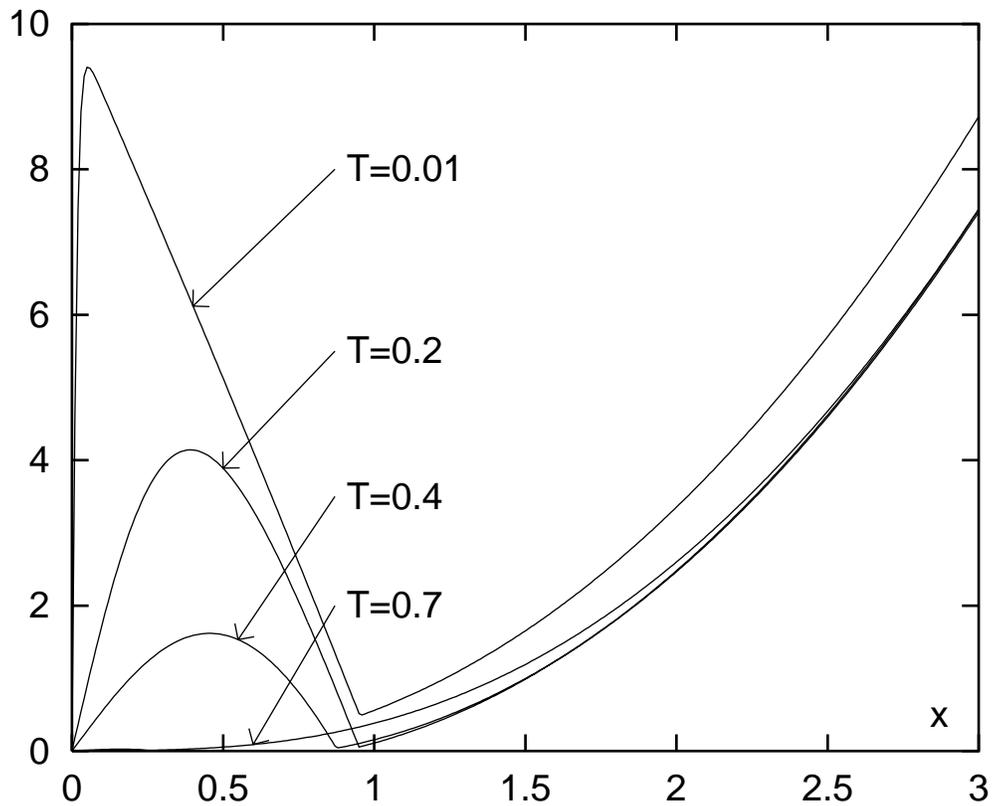}
\caption{The left hand side of the gap equation (which is proportional
to the derivative of the effective potential) as the function of
$x=\sigma/\mu$. The input parameters are  $\Lambda=10\mu$,
$a_{\sigma}^{(0)}=a_{\pi}^{(0)} =5/\mu$, and $N=50$.} 
\label{Figure4}
\end{figure}

\end{document}